\documentclass[sigconf, nonacm]{acmart}

\usepackage{lineno}

\AtBeginDocument{%
  \providecommand\BibTeX{{%
    \normalfont B\kern-0.5em{\scshape i\kern-0.25em b}\kern-0.8em\TeX}}}

\begin{document}

\title{A RISC-V SystemC-TLM simulator}

\author{Marius Monton}
\email{marius.monton@uab.cat}
\orcid{}
\affiliation{%
  \institution{Departament de microelectrònica i sistemes electrònics \\ Universitat Autònoma de Barcelona}
  \city{Barcelona}
  \state{Spain}
  \postcode{08193}
}

\renewcommand{\shortauthors}{M. Montón}

\begin{abstract}
    This work presents a SystemC-TLM based simulator for a RISC-V microcontroller. 
    This simulator is focused on simplicity and easy expandable of a RISC-V. It is built around a full RISC-V instruction set simulator that supports full RISC-V ISA and extensions M, A, C, Zicsr and Zifencei. 
    
    The ISS is encapsulated in a TLM-2 wrapper that enables it to communicate with any other TLM-2 compatible module. 
    The simulator also includes a very basic set of peripherals to enable a complete SoC simulator. The running code can be compiled with standard tools and using standard C libraries without modifications. 
    
    The simulator is able to correctly execute the riscv-compliance suite. The entire simulator is published as a docker image to ease its installation and use by developers. A porting of FreeRTOSv10.2.1 for the simulated SoC is also published.
\end{abstract}



\maketitle

\section{Introduction}
Many simulators has been published since the release of first drafts of RISC-V ISA \cite{RISCVsim}. These simulators use different techniques and technologies to achieve different requirements: good performance, good visualization of the processor, architectural exploration, etc. Most of them conform to RISC-V ISA specifications; some of them use a previous infrastructure and adapt the ISS to follow the RISC-V ISA and re-uses some peripherals already simulated \cite{QEMU, Imperas, gem5, ta2018simulating}. Others are written from scratch and includes the ISS and a minimum set of peripherals \cite{Spike, mallya2018flexible}. There are FPGA-based simulators to increase performance and simulation speed \cite{Rocket} as well as the precision of the simulation results.

The Spike simulator is most common simulator and it is used as reference model for RISC-V ISA \cite{Spike}. Other simulators are intended for a graphical visualization for the entire execution of the instructions inside the CPU \cite{ripes}.


SystemC is a set of libraries for the C++ language to allow the description and simulation of hardware based systems by a event-driven simulation model. This libraries add time management, concurrency and hardware-like data types to C++ \cite{SystemC}.

Transaction Level Modelling adds a layer to SystemC in order to model the interface between different modules in a lightweight way. This model technique uses transactions to abstract any kind of communication between modules, hiding or avoiding the details of the communication itself: a transaction is an access from a Master (called Initiator) to a Slave (called Target) to a memory address with a length and some attributes. The Slave will respond to the transaction within a time (that can be 0 for a basic modeling) and the writing or reading of the transaction. All other details of the transaction (bus access, signals change, etc.) are not modeled. In more detailed modeling, the different phases of a bus access can be specified. Currently, SystemC standard includes TLM modeling \cite{SystemC}. The modules can also interchange data using direct pointers to memory instead to transactions to increase simulation speed. This technique is named Direct Memory Interface (DMI).

TLM has boosted the interoperability between vendors models and the appearance of many IPs that are interchangeable and fully compatible among different systems and vendors. The fundamental idea of this work is to introduce all these features to a RISC-V simulator.

The source code of the entire project is open-source and published \cite{github}. 

The presented simulator is intended for an easy use and simple to extend, with clear code and able to simulate an entire SoC, like any embedded microcontroller in the market. To keep the code simple, meta-programming has been avoided and C++ templates use is keep as low as possible.

The paper is structured in the following sections: Section II depicts the architecture of the entire simulator, Section III show software particularities and tool-chain modifications, Section IV shows simulation performance and compliance results. Section V concludes the paper.

\section{Simulator Architecture}


One of the main goals of this simulator was to be easily extensible and modifiable. To achieve this objective, the original design was very simple and clear, with the use of naive techniques and a source code designed for simplicity.

\begin{figure}
    \centering
    \includegraphics[width=\linewidth]{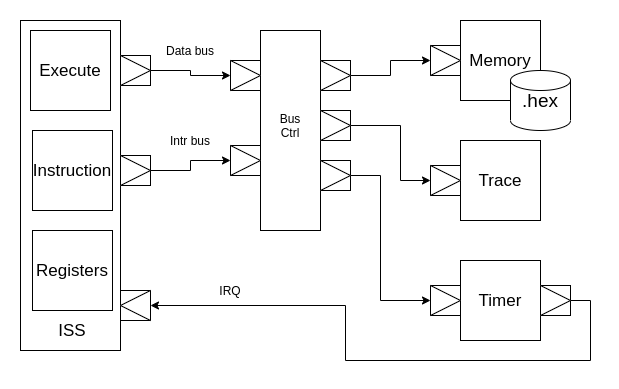}
    \caption{TLM Diagram of the entire simulator}
    \label{fig:TLMDiagram}
\end{figure}

The simulator architecture includes a ISS for RV32I ISA \cite{riscv-isa}, a bus controller, the main memory and peripherals. Communication between these modules is done by TLM-2 sockets (see Figure~\ref{fig:TLMDiagram}).

\subsection{CPU}

The ISS simulates a single hardware thread (HART) and includes privileged instructions. It is divided in three modules: {\it Instruction}, {\it Execute} and {\it Registers}:
\begin{itemize}
    \item {\it Instruction} Decodes instructions and checks for extensions. This module can access all fields of each instruction type (R, I, S, B, U and J type).
    \item {\it Execute} Executes instructions, accessing registers and memory and performing operations. This module also executes "MACZicsr\_Zifencei" extensions \cite{riscv-isa}. 
    \item {\it Registers} Implements the register file for the entire CPU, including general-purpose registers (r0-r31), Program counter (pc) and all necessary entries in Control and Status Registers (CSR) registers.
\end{itemize}

This CPU is a minimal, fully functional model with a end-less loop fetching and executing instructions without pipeline, branch predictions or any other optimization technique. All instructions are executed in one single cycle, but it can be easy customized to per instruction cycle count.

The {\it Execute} module implements each instruction with a class method that receives the instruction register. These methods perform all necessary steps to execute the instruction. In case of a branch instruction, these methods are able to change the PC value. For Load/Store instructions, the methods are in charge to access the required memory address. 

The CPU is designed following Harvard architecture, hence the ISS has separate TLM sockets to interface with external modules:
\begin{itemize}
    \item Data bus: Simple initiator socket to access data memory.
    \item Instruction bus: Simple initiator socket to access instruction memory.
    \item IRQ line: Simple target socket to signal external IRQs.
\end{itemize}

\subsection{Bus Controller}
The simulator also includes a Bus controller in charge of the interconnection of all modules. The bus controller decodes the accesses address and does the communication to the proper module. In the actual status of the project, it contains two target sockets (instruction and data buses) and three initiator sockets: {\it Memory}, {\it Trace} and {\it Timer} modules, as described below. 

\subsection{Peripherals}

The {\it Memory} module simulates a simple RAM memory, which is the main memory of the SoC, acting as instruction memory and data memory. This module can read a binary file in Intel HEX format obtained from the .elf file and load it to be the main program for the ISS. This module has a Simple target socket to be accessed that supports DMI to increase simulation speed.

The simulated Soc includes a very basic {\it Timer} module. This module includes two 64 bits register mapped to 4 addresses. On of this registers (\textit{mtime}) keeps current simulated time in nanosecodns resolution. The second register (\textit{mtimecmp}) is intended to program a future IRQ. The module triggers an IRQ using its Simple initiator socket. 

\begin{figure}
    \centering
    \includegraphics[width=\linewidth]{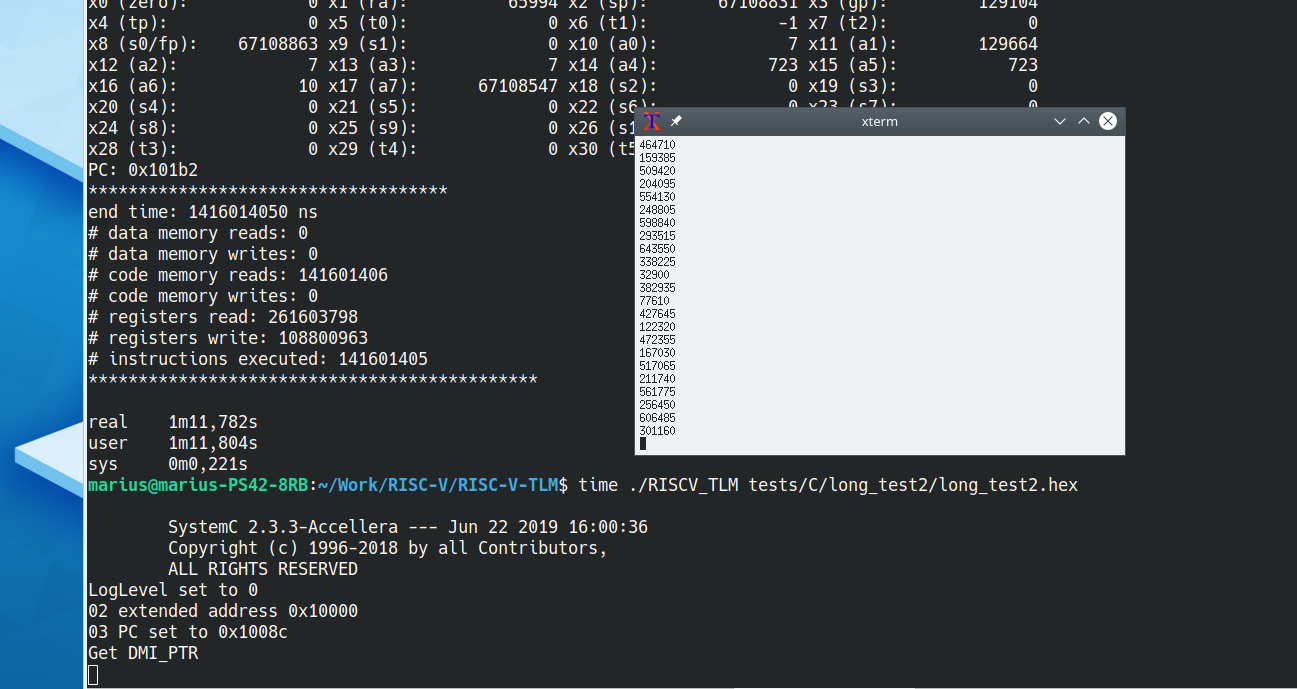}
    \caption{Simulator running with an xterm windows as terminal}
    \label{fig:RISCVTLMScreenshot}
\end{figure}

The {\it Trace} module is a very simple tracing device, that outputs through a xterm window the characters received. This module is intended as a basic mimic of the ITM module of Cortex-M CPUs \cite{ITM}.  Figure \ref{fig:RISCVTLMScreenshot} shows the simulator running with an xterm windows as output console.

\begin{figure}
    \centering
    \includegraphics[width=\linewidth]{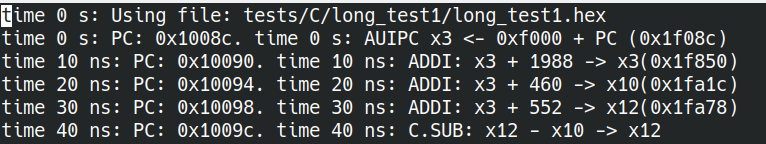}
    \caption{Log file view}
    \label{fig:logfile}
\end{figure}

Two other modules are included in the simulator: {\it Performance} and {\it Log}. The {\it Performance} module take statistics of the simulation, like instructions executed, registers accessed, memory accesses, etc. It dumps this information when the simulation ends.
The other module allows the simulator to create a log file with different levels of information.\\
At maximum level of logging, each instruction executed is logged into the file with its name, address, time and register values or addresses accessed.
The log file at maximum debug level shows information about the current time, PC value and the instruction executed.
It also prints the values of the registers used. Figure \ref{fig:logfile} shows a real executed log file.


The log file at maximum debug level shows information about the current time, PC value and the instruction executed. It also prints the values of the registers used. Figure~\ref{fig:logfile} shows a real executed log file.

\section{Software implementation and toolchain}

The entire simulator is designed to work on pure bare-metal simulation. There is not direct communication between the simulator and the host machine, meaning for instance that \textit{printf} implementation outputs directly to a host computer console. This is intended to do a simulation as similar to a real Hardware as possible, because the same exact code and the compiled binary that runs in the simulator will run in the real SoC.

For this reason the instructions {\it EBREAK} and {\it ECALL} are implemented in that way: {\it EBREAK} stops the simulation and dump some statistics. In a real system, has no sense to call \textit{EBREAK} instruction and depending of the implementation can trigger a system reset or a \textit{NOP}.  The {\it ECALL} instruction raises an exception, dump statistics of the simulation and continues the execution for the same reason.

To supply the lack of semi-hosting options, the {\it Trace} module can be used to print out some information. With the use of proper helper functions, it is possible to use {\it printf() }-like functions. In this case, the {\it \_write} function must be written to send the received data to {\it Trace} module as follows: 

\begin{verbatim}
  int _write(int file, const char *ptr, int len) 
  {
    int x;

    for (x = 0; x < len; x++) {
      TRACE =  *ptr++;
    }

    return (len);
  }
  
\end{verbatim}
\label{code:write}

The initial value for the Program Counter register (PC) is obtained from the HEX binary file and set before starting the simulation. The stack pointer register (SP) is set to last memory address.  

This flexibility and the compatibility accomplished enables the use of the standard GCC cross compiler with little options:
\begin{verbatim}
    -march=rv32imac -mabi=ilp32 --specs=nosys.specs
\end{verbatim}

The options specifies the architecture and ABI (Application Binary Interface) and specifies the bare-metal option for newlib standard C library.

This allows complete use of C library on the application code, including math library, stdio and string libraries.

\subsection{Docker version}
A docker version of the simulator is provided \cite{Docker}. It can be used to ease the installation and use of the simulator to avoid user to compile and gather all necessary libraries. 

This image has been used in conjunction with another docker image that contains a riscv-toolchain. It can be used to ease the installation and use of the simulator, and specifically, to avoid the user to compile and gather all necessary libraries.

The simulator image is published and available in docker hub \cite{Dockerhub}.

\subsection{FreeRTOS}

A porting of FreeRTOS version 10.2.1 were written for the simulated SoC \cite{FreeRTOS}. The simulator is able to run this complex project without any error. 
The  FreeRTOS test project includes 3 tasks that communicate and synchronize using one common queue. The two producer tasks use FreeRTOS' delay functions to suspend for a amount of time. Only one of the tasks prints out debug information.

\section{Test and Results}

Different test were done to ensure the compatibility of the simulator. Also some performance results are presented from the same tests.

The compiler for RISC-V code is the RISC-V GCC version is 8.3.0 build with ABI configured to {\it ilp32} and architecture set to {\it rv32i}.

\subsection{Tests Compliance}

The simulator implements RISC-V RV32IMACZicsr\_Zifencei V2.1 instruction set \cite{riscv-isa, riscv-isa_p} and it passes all tests in risc-test and riscv-compliance suites \cite{RISCVTest, RISCVCompilance}. The riscv-compliance tests have a coverage of 97.23\% for RV32I, 89.95\% for RV32IM and 59.68\% for RV32IMC. These percentage means the number of all possible instructions and registers combinations are tested.

A more complex program, the {\it dhrystone} benchmark test is passed with correct results as well. 

The project code has been statically checked with {\it coverity} by Synopsis. The analysis results in only 1 minor error found in TLM-2 library code but any error in the simulator code itself \cite{Coverity}. Also, code quality is checked with {\it Codacy} tool \cite{Codacy}. This tool checks for code quality, security, unused code, etc. The outcome of this tool is a A score, with only 10 minor warnings about code style.

In the next section is discussed the performance of this simulator.

\subsection{Performance}

A set of four program are written to test the performance of the simulator. Of these tests, test 1 checks memory transfer between two memory locations; test 2 and 3 perform arithmetic operations in three variables, one prints out the results and the other one is not using the console; the last test uses string manipulation functions from stdlib C library (printf, sprintf, strcpy). 

\begin{table}
  \caption{Performance result. Values in instructions/second}
  \label{tab:performance}
  \begin{tabular}{ccl}
    \toprule
    Test & Native & Docker \\
    \midrule
    Test1 & 8.252.929 & 3.854.110 \\
    Test2 & 6.298.774 & 3.291.465 \\
    Test3 & 8.921.763 & 3.754.295 \\
    Test4 & 12.899.367 & 4.375.651 \\
    Dhrystone & 10.700.733 & 3.796.328 \\
  \bottomrule
\end{tabular}
\end{table}

\begin{figure}
    \centering
    \includegraphics[width=\linewidth]{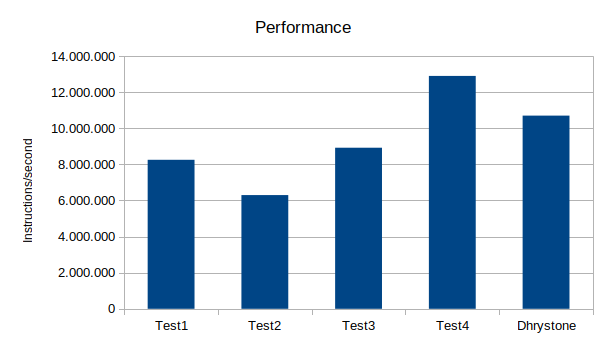}
    \caption{Execution results for all tests}
    \label{fig:performance}
\end{figure}

All test do a end-less loop of some mathematical operations and prints out the result using {\it Trace} module. Each test is executed 3 times for different execution time (from 10 to aprox. 60 seconds execution time). The Figure~\ref{fig:performance} shows average of these 3 runs.

Its performance varies mainly with the level of the logging system due to huge I/O traffic in the log file. With lowest level o logging, the performance of the simulator is about 8 million of simulated instructions per second (see Table~\ref{tab:performance} and Figure~\ref{fig:performance}) in a Intel Core i7-8550U CPU @ 1.88 GHz with 16 GB of memory. As a reference, in the same computer the Spike simulator performance is about 170 million of simulated instructions per second.

The low performance of Test2 can be due to the intensive use of the {\it Trace} module and the overhead it implies.

For the {\it Dhrystone} benchmark, it is executed with good results and the performance is about 7200 Dhrystones/second. It has been tested with 10.000, 250.000 and 500.000 loops of the Dhrystone test.

\subsection{Docker version}

\begin{figure}
    \centering
    \includegraphics[width=\linewidth]{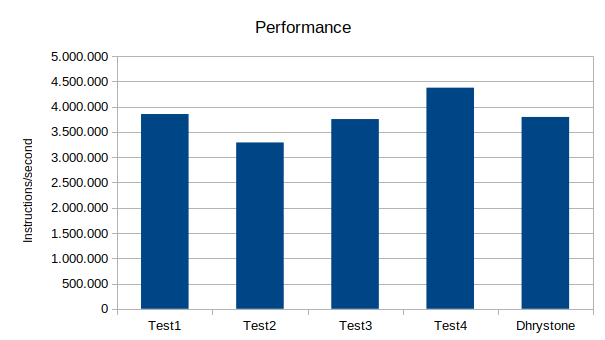}
    \caption{Execution results for all tests with the Docker version}
    \label{fig:performanceDocker}
\end{figure}

The same tests has been run with the docker version of the simulator. The results are summarized in Table~\ref{tab:performance} and depicted in Figure~\ref{fig:performanceDocker}.

In case of docker version, the performance has a penalty from 47\% to up to 69\% depending on the test. The performance of this version is depicted in Figure~\ref{fig:performanceDocker}.

\newpage

\section{Conclusions}
This paper introduces a new RISC-V simulator. It has been designed from scratch to simulate an entire SoC with simplicity on focus. It has been designed in SystemC and TLM-2 as language and modeling schema.

It has been presented the main architecture of the simulator, the software configuration and tools required. Followed by a brief discussion about the simulation performance and the conformance to the specifications.

The use of standards is important in any aspects of the engineering effort. In the case of system-level simulators, the existence of the TLM-2 and SystemC standards should be encourage and used by vendors and researchers to increase the interoperability and re-usability of the components. This simple simulator is a first step towards this achievement.


\bibliographystyle{ACM-Reference-Format}
\bibliography{CARRV}

\end{document}